\documentclass{article}
\usepackage{amsmath}
\usepackage{multicol}

\usepackage{mathptmx}           
\usepackage{graphicx}           
\usepackage{url}                
\usepackage{float}
\bibliography{refs}
\usepackage{fullpage}
\usepackage{capt-of}

\usepackage{textcomp}
\usepackage[english]{babel}

\usepackage{xspace}

\usepackage[none]{hyphenat}

\begin{document}

\author{Matthew Long, Aditya Jami, Ashutosh Saxena\\
  Stanford University}
\title{Hierarchical classification of e-commerce related social media}
\date{}
\maketitle

\begin{multicols}{2}
\begin{abstract}
In this paper, we attempt to classify tweets into root categories of the Amazon browse node hierarchy using a set of tweets with browse node ID labels, a much larger set of tweets without labels, and a set of Amazon reviews. Examining twitter data presents unique challenges in that the samples are short (under 140 characters) and often contain misspellings or abbreviations that are trivial for a human to decipher but difficult for a computer to parse. A variety of query and document expansion techniques are implemented in an effort to improve information retrieval to modest success.
 \end{abstract}

\section{Introduction}
Internet users post information regarding a topic on a number of different websites, but companies and organizations typically only train their classification algorithms using only the information posted on their own platform. Obviously, data from competitors is often difficult to acquire, but in cases where it is freely available, cross-platform analysis can only benefit a model as data from other sources can be used only if it improves performance. In order for this data to be valuable, it has to be correctly classified by what it refers to. 

The goal of this project is to to find a likely product category within the root categories of the Amazon browse node hierarchy for a given tweet. Twitter data consisted of a training dataset with 58,000 tweets labeled with Amazon browse node IDs, and a much larger set of 15,000,000 unlabeled tweets that can be used for augmentation. The Amazon data consisted of 1,900,000 reviews for products labeled by their browse node ID. All of the datasets originally were in JSON format and contained metadata as well as text content for each review or tweet. To obtain root nodes for tweets, a browse node ID tree was created so that a simple parent traversal could identify a root category. The Amazon product hierarchy is more of a directed graph in that it children categories can have multiple parents. In these cases, the parent is chosen randomly. 28 root categories were identified from the browse nodes within the labeled dataset, but the distribution was heavily skewed, with 47,000 tweets in the \textit{books} root category, and 10 or fewer in 5 categories. Furthermore, over half of the tweets were re-tweets, which have the same text content as the original tweet, providing no additional information to a classifier while inflating accuracy misleadingly. Once re-tweets and tweets from categories with fewer than 5 tweets were removed, the labeled corpus contained 23,910 tweets from 24 root categories.

\begin{center}
 \footnotesize
 \label{table:top5}
    \begin{tabular}{| l | l | p{3cm} |}
    \hline
    {\it Amazon Category} & {\it Tweets} & {\it Common Keywords} \\ \hline
    Books & 19531 & 'free', 'ebook', 'kindle' \\ \hline
    Home \& Kitchen & 1332 & 'clock', 'lp', 'wall' \\ \hline
    Clothing, Shoes \& Jewelry & 647 & 'woman', 'dress', 'skirt' \\ \hline
    Movies \& TV & 643 & 'dvd', 'video', 'instant' \\ \hline
    Electronics & 403 & 'apple', 'hd', 'sony'  \\ \hline
    \hline
    \end{tabular}
    \captionof{table}{Top 5 categories by number of tweets}
\end{center}

\section{Method}
\subsection{Preprocessing}
As the data consists of text strings, a bag-of-words model was used to represent the tweets. To reduce feature size and trim unhelpful data, all the tweets were converted to lower case and stripped of all punctuation except hashtags. Additionally, URLs and stop words from both a list within the Natural Language Toolkit and a list we developed specifically for Twitter were removed and words were stemmed with the WordNet Lemmatizer \cite{nltk}\cite{wn}. With 5-fold cross validation, corresponding to an 80/20 training/testing split, the unprocessed tweets had 48,000 unique words, which got truncated to 22,213 words after pre-processing. Text was then transformed to a sparse matrix representation of TF-IDF features in order to be acceptable for downstream estimators. This weighting scheme was chosen because it weights against words that show up frequently across all documents and thus implicitly reflects the importance of a word in a document \cite{Tfidf}\cite{term}. TF-IDF refers to the term-frequency multiplied by the inverse document frequency and is calculated as,
\[ \mathit{tf_{ij}} = \frac{f}{\sum\limits_{i}f_{ij}}\]
\[ \mathit{idf_{i}} = \log \frac{N}{\mathit{df_{i}}} \]

\subsection{Baseline Models}
To evaluate the impact of our tests, we compared different learning algorithms performance when trained on the preprocessed dataset with all features. To ensure that there were both training and testing examples for each category a stratified 5-fold cross-validation was used to split up the dataset into training and testing sets. The metrics associated with each classifier indicate the unweighted mean of the metrics for each category. We choose to evaluate model quality in this fashion because of the imbalanced nature of the labeled dataset. The vectorization of the corpus and the training of the models were done using the Scikit-Learn package\cite{scikit-learn}.
\begin{center}
 \footnotesize
 \label{table:bc}
    \begin{tabular}{| l | l | l | l |}
    \hline
    {\it Classifier} & {\it Precision} & {\it Recall} & {\it F1 Score} \\ \hline
    Multinomial NB & 9.2\% & 23.6\% & 0.10  \\ \hline
    Logistic Regression & 46.1\% & 73.7\% & 0.54 \ \\ \hline
    Linear SVM & 48.0\% & 75.6\% & 0.56 \\ \hline
    Linear SVM (w/ class weights) & 51.7\% & 72.3\% & 0.58 \\ \hline
    \hline
    \end{tabular}
    \captionof{table}{Baseline Classifier Average Test Scores}
\end{center}

Class weights provide a way of correcting for imbalance by weighting for or against certain classes but would be difficult to tune for each technique we will explore\cite{bal}. For this reason, an unweighted linear SVM will be used as the baseline against which to measure the effectiveness of our approach, although class weights will be used for final model. The evaluation metric for these comparisons will be the F1-score, as it combines precision and recall into a single number.  

\[ \mathit{F_{1}} = 2\cdot\frac{\mathit{precision}\cdot\mathit{recall}}{\mathit{precision}+\mathit{recall}}\]

\subsection{Feature Selection}
Features were ranked according to their Anova F-values and models were evaluated when trained on the top \textit{n} percent of features\cite{feature}. We trained models for unigram features and unigram and bigram features.

\begin{figure}[H]
\includegraphics[width=\columnwidth]{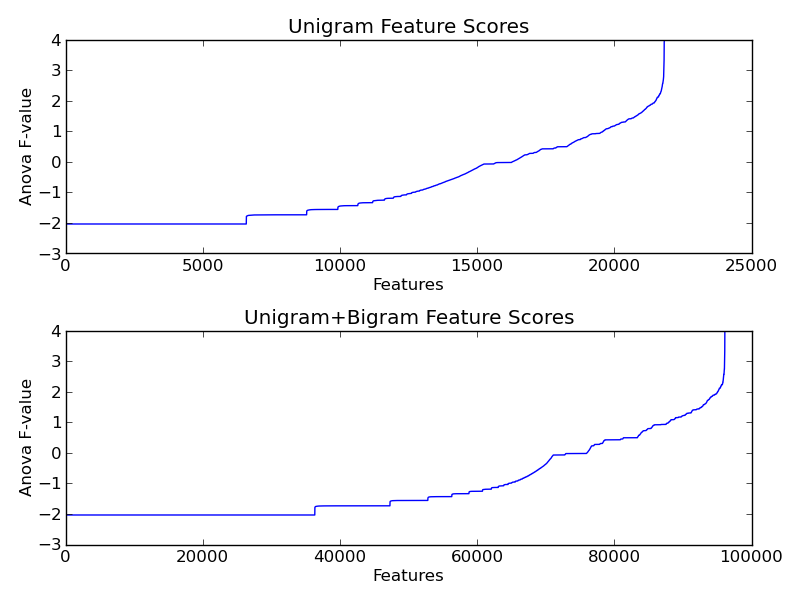}
 \caption{Anova F-values for unigram features and unigram+bigram features}
 \label{fig:Fscore}
\end{figure}

\begin{figure}[H]
\includegraphics[width=\columnwidth]{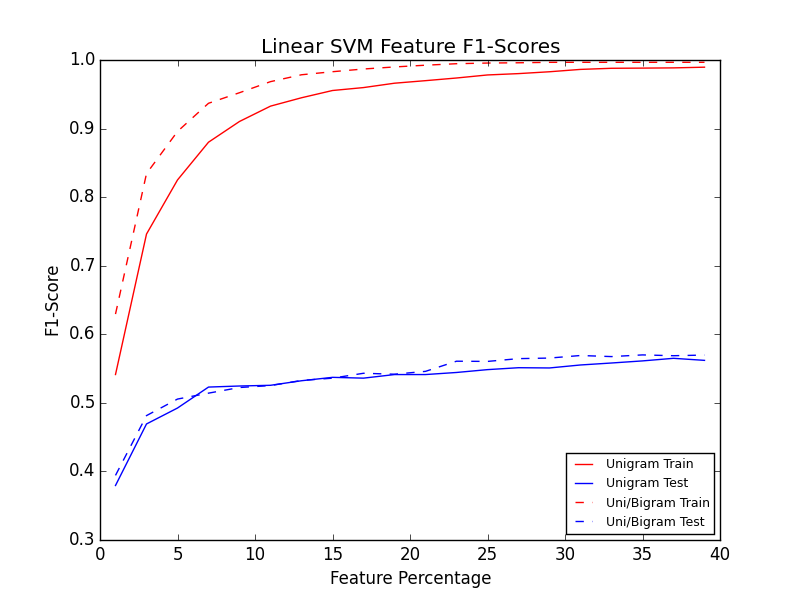}
 \caption{F1-scores for unigram features and unigram+bigram features}
 \label{fig:FeatScore}
\end{figure}

It is clear from Figure \ref{fig:FeatScore}, that precision and recall in the test set stabilize after using around 20\% of the features in both the unigram and unigram+bigram cases.  As the F1-score for both of these cases were roughly similar, and the absolute number of features for a given percentage is much lower for only unigram features, we decided to use 25\% of the unigram features for our models.

\subsection{Expansion}
As tweets are shorter than typical documents, expanding them seems reasonable as it improves the vocabulary of the model\cite{Exp}. In order to improve classification accuracy, we considered query expansion, in which terms are added to testing tweets, and document expansion, in which terms are added to training tweets. Both topics are areas of research in Information Retrieval (IR), although query expansion is the more promising, and thus more studied field\cite{micro}.

\subsubsection{Document}
Tweets from the training set were expanded based upon hashtags contained within them and the root category they belonged to. To perform hashtag expansion a thesaurus was built up of the most frequent words in tweets containing a given hashtag using the large unlabeled Twitter dataset. \textit{n}  randomly selected words from the top 2\textit{n} words from each hashtag were then added to each tweet containing that hashtag. No words from the stop lists would be added, nor would the hashtag word. For root category expansion, one thesaurus was built using for each category using the words in the training set portion of the labeled tweets and another was built for the reviews in the Amazon set. When building the thesaurus for root category expansion using Twitter, the top words for each category were chosen with a TF-IDF weighting scheme, however, because the corpus the thesaurus was built upon was much smaller allowing the process to be computationally feasible.
\subsubsection{Query}
As the hashtag thesaurus was built from an external dataset, hashtag expansion could be used on tweets from the testing set portion of the labeled tweets as well. An identical process to document hashtag expansion was used.

\begin{center}
 \footnotesize
 \label{table:bc}
    \begin{tabular}{| p{3.5cm} | p{3.5cm} |}
    \hline
    {\it Tweet} & {\it Suggested Expansion Words} \\ \hline
    "wepnewlyrelease new read bulletin board \#fiction \#thriller" & 'review', 'book', 'fiction', 'literature', 'child' \ \\ \hline
    "aburke59 dead sister jessica huntington deser free sample \#mystery" & 'get', 'read', 'star', 'murder' \ \\ \hline
    \hline
    \end{tabular}
    \captionof{table}{Hashtag Expansion Examples}
\end{center}

\section{Results}
\subsection{Expansion}
\begin{figure}[H]
\includegraphics[width=\columnwidth]{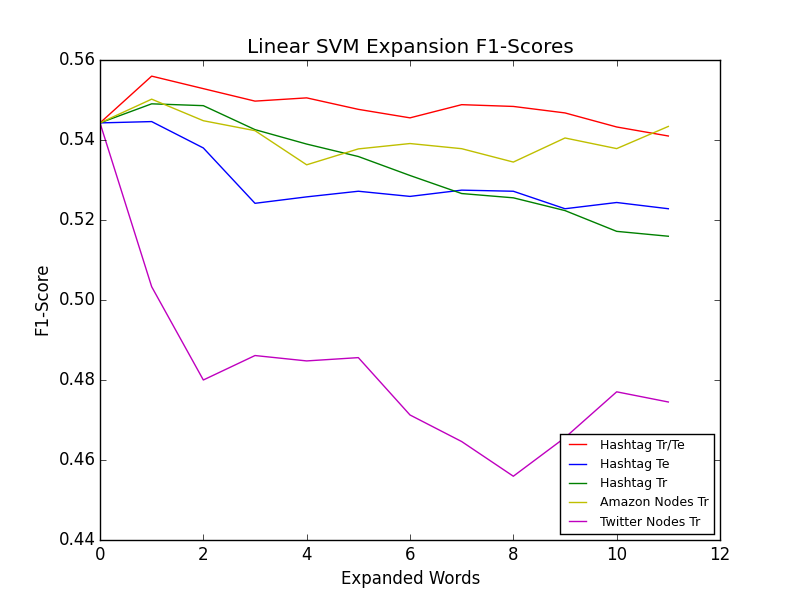}
 \caption{F1-scores for various expansion techniques}
 \label{fig:ExpScore}
\end{figure}

Tweet expansion saw mixed results in category classification. Hashtag expansion on both the training and testing set marginally improved performance, while hashtag expansion on each set exclusively worsened performance. Amazon node expansion achieved similar results as the base case model, while Twitter node expansion significantly decreased performance. Figure \ref{fig:ExpScore} details the results expansion for various expansion lengths.

\subsection{Overall}
In the final model, we used both hashtag document and query expansion and also added class weights to the linear SVM classifier. The class weighting scheme that was added was primarily directed at reducing the effects of the imbalance toward the \textit{books} category so a weight of 0.1 was applied to that category, while other categories weighted by 1\cite{bal}. Additionally, the C parameter of the SVM estimator was tuned using the GridSearch function of Scikit-Learn and a value of 5 was selected. Table 4 shows the results of our final model.

\begin{figure}[H]
\includegraphics[width=\columnwidth]{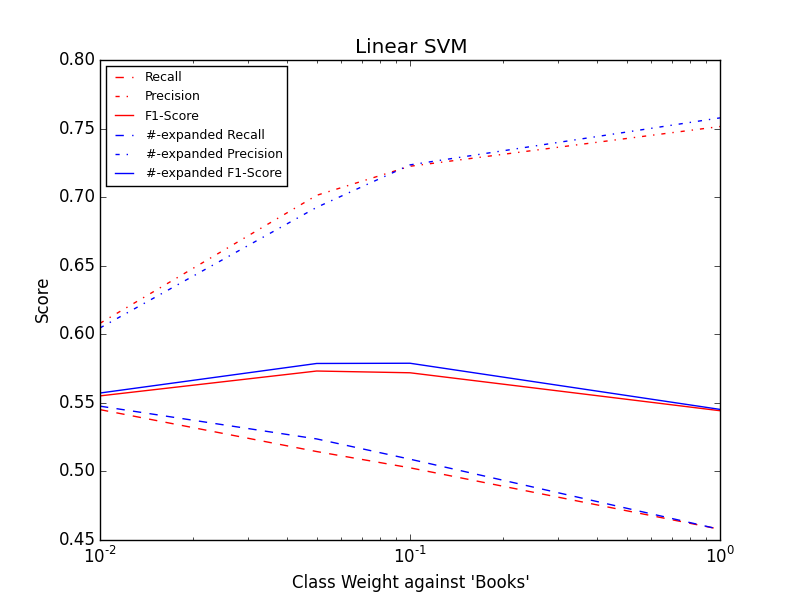}
 \caption{Scores for various class weights against books}
 \label{fig:CWScore}
\end{figure}

\begin{center}
 \label{table:resu}
 {\scriptsize
    \begin{tabular}{| p{2.25cm} | l | l | l | l |}
    \hline
    {\it Amazon Category} & {\it Precision} & {\it Recall} & {\it F1-Score} & {\it Support}\\ \hline
    Baby Products & 0.89 & 1.00 & 0.94 & 8\\ \hline
    Health \& Personal Care & 0.78 & 0.85 & 0.81 & 46\\ \hline
    Digital Music & 0.82 & 0.64 & 0.72 & 22\\ \hline
    Beauty & 0.71 & 0.50 & 0.59 & 10\\ \hline
    Sports \& Outdoors & 0.69 & 0.62 & 0.65 & 56\\ \hline
    Arts, Crafts \& Sewing & 1.00 & 0.20 & 0.33 & 5\\ \hline
    Video Games & 0.89 & 0.53 & 0.67 & 32\\ \hline
    Home \& Kitchen & 0.84 & 0.89 & 0.87 & 334\\ \hline
    Kindle Store & 1.00 & 0.67 & 0.80 & 3\\ \hline
    Tools \& Home Improvement & 0.75 & 0.50 & 0.60 & 18\\ \hline
    Collectibles \& Fine Art & 0.87 & 0.81 & 0.84 & 16\\ \hline
    CDs \& Vinyl & 0.83 & 0.35 & 0.49 & 55\\ \hline
    Patio, Lawn \& Garden & 0.00 & 0.00 & 0.00 & 7\\ \hline
    Clothing, Shoes \& Jewelry & 0.89 & 0.76 & 0.82 & 162\\ \hline
    Cell Phones \& Accessories & 1.00 & 0.14 & 0.25 & 7\\ \hline
    Books & 0.96 & 0.98 & 0.98 & 4883\\ \hline
    Pet Supplies & 1.00 & 0.11 & 0.20 & 9\\ \hline
    Automotive & 1.00 & 0.60 & 0.75 & 5\\ \hline
    Musical Instruments & 1.00 & 0.70 & 0.82 & 10\\ \hline
    Movies \& TV & 0.74 & 0.69 & 0.71 & 161\\ \hline
    Office Products & 1.00 & 0.56 & 0.71 & 9\\ \hline
    Toys \& Games & 0.86 & 0.24 & 0.38 & 25\\ \hline
    Electronics & 0.82 & 0.75 & 0.78 & 101\\ \hline
    Grocery \& Gourmet Food & 0.00 & 0.00 & 0.00 & 2\\ \hline
    {\it Category Average} & 0.81 & 0.54 & 0.61 & 5896\\ \hline
    {\it Absolute Average} & 0.94 & 0.94 & 0.94 & 5896\\ \hline
    \hline
    \end{tabular}
    }
    \captionof{table}{Model Results for 75/25 training/testing split}
\end{center}

\section{Discussion}
\begin{figure}[H]
\includegraphics[width=\columnwidth]{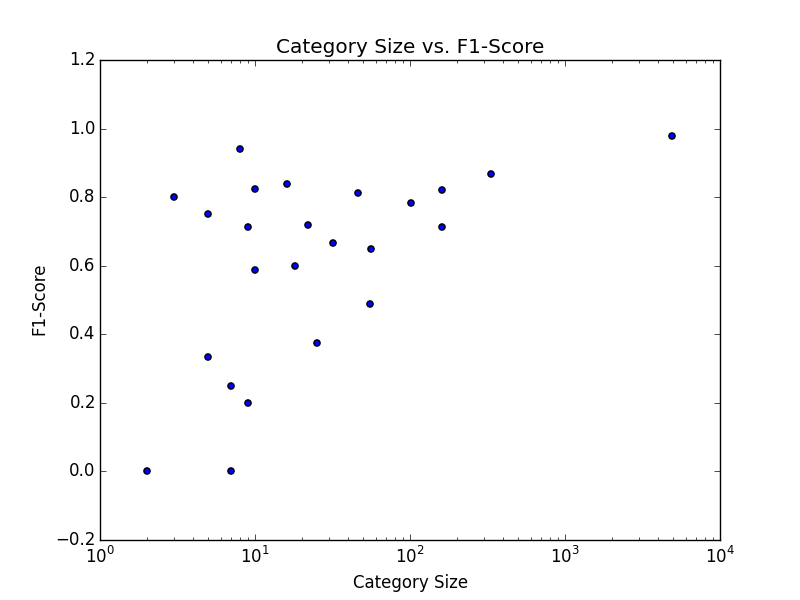}
 \caption{Category Size compared with F1-Scores}
 \label{fig:CatSize}
\end{figure}

The model achieved an average F1-score across all categories of 0.61 with average precision of 81\% and average recall of 54\%. Categories with more tweets tended to be classified more accurately than tweets with few samples to draw upon. This makes intuitive sense as the vocabulary of the samples in the small categories is limited so there are high odds that the testing samples do not contain the same words as in the training samples. This is representative of the fact that the bound on generalization error decreases as the sample size increases, so naturally larger categories are capable of better testing accuracy. Figure \ref{fig:CatSize} demonstrates this rough trend. Query expansion is typically regarded to be more effective than document expansion and the only thing we expanded in the test set were hashtags\cite{micro}. Many tweets do not contain any hashtags, so the effects of query expansion was only received by a fraction of the test set. It is clear that using external datasets (ie. Amazon, unlabeled twitter) to augment the labeled twitter set do not decrease performance. Less clear, however, is whether these dataset can be better leveraged to significantly improve performance.

\section{Future Work}
The next step to take would be to build up a thesaurus on individual words from both Amazon and unlabeled Twitter data in order to expand testing and training tweets on a per word basis. Building these thesauruses will be space intensive because for each word the frequency of all the other words it has appeared with in a tweet or review has to be stored. This step holds promise as it could be used for both query and document expansion and could be used upon all tweets. With a full word thesaurus, selective expansion could also be explored, where only certain categories are expanded. There are existing thesauruses that can be downloaded such as WordNet, but the frequent use of abbreviations and slang on Twitter makes building a thesaurus from a corpus of tweets potentially more beneficial\cite{wn}.  Another step that would provide immediate benefits is building a larger corpus for under-represented categories. An alternative to hand labeling additional tweets would be to make use of semi-supervised learning techniques that can leverage the large unlabeled dataset to improve performance.


\end{multicols}

\end{document}